\documentclass{aa}  
\usepackage{natbib}
\bibpunct{(}{)}{;}{a}{}{,}

\usepackage{graphicx}
\usepackage{txfonts}

\newenvironment{myfont}{\fontfamily{pcr}\selectfont}{\par}
\DeclareTextFontCommand{\textmyfont}{\myfont}
\def\gtsima{$\; \buildrel > \over \sim \;$} 
\def\ltsima{$\; \buildrel < \over \sim \;$}  
\def\gsim{\lower.5ex\hbox{\gtsima}}
\def\simgt{\lower.5ex\hbox{\gtsima}} 
\def\simlt{\lower.5ex\hbox{\ltsima}} 
\def\lsim{\lower.5ex\hbox{\ltsima}}

\usepackage[]{hyperref}
\begin{document}

   \title{X-ray absorbing column densities of a complete sample of short Gamma Ray Bursts}

   \author{L. Asquini
          \inst{1,2}
%          \inst{3}
          \and
          S. Campana\inst{2}
          \and
          P. D'Avanzo\inst{2}          
          \and 
          M.G. Bernardini\inst{4,2}
          \and
          S. Covino\inst{2}
          \and 
          G. Ghirlanda\inst{2,3}
          \and
          G. Ghisellini\inst{2}
          \and
          A. Melandri\inst{2}
          \and
          L. Nava\inst{2,5,6}
          \and
          O. S. Salafia\inst{2,3,7}
          \and
          R. Salvaterra\inst{8}
          \and
          B. Sbarufatti\inst{9}
          \and
          G. Tagliaferri\inst{2}
          \and
          S. D. Vergani\inst{10,2}
		}
   \institute{Dipartimento di Fisica, Universit\`a degli Studi di Milano, Via Celoria 16, I-20133 Milano, Italy\\
   				\email{laura.asquini@studenti.unimi.it}
		 \and
             INAF-Osservatorio Astronomico di Brera, Via Bianchi 46, I-23807 Merate (LC), Italy
             \and
             Universit\`a degli Studi di Milano-Bicocca, Piazza della Scienza 3, I-20126 Milano, Italy               
%             \email{sergio.campana@brera.inaf.it} \\
%             \email{paolo.davanzo@brera.inaf.it}
		\and 
		    Laboratoire Univers et Particules de Montpellier, Universit\'e Montpellier, CNRS/IN2P3, Montpellier, France		
		\and
		    INAF-Osservatorio Astronomico di Trieste, Via G. B. Tiepolo 11, I-34143, Trieste, Italy
		\and
		    INFN-Istituto Nazionale di Fisica Nucleare, Sezione di Trieste, Via Valerio 2, I-34127, Trieste, Italy
		\and 
		    INFN-Sezione di Milano-Bicocca, Piazza della Scienza 3, I-20126 Milano, Italy		    
		\and
		    INAF-Istituto di Astrofisica Spaziale e Fisica Cosmica Milano, Via E. Bassini 15, I-20133 Milano, Italy
		\and
		    Department of Astronomy and Astrophysics, The Pennsylvania State University, University Park, PA 16802, USA 
		\and
		    GEPI, Observatoire de Paris, PSL Research University, CNRS, Place Jules Janssen, F-92195 Meudon, France		
	  }

%   \date{data}

% \abstract{}{}{}{}{} 
% 5 {} token are mandatory
 
  \abstract
   {X-ray absorbing column densities ($N_{H}$) are used as a parameter to quantify the amount of absorbing material along the line of sight. The high values found for long Gamma-Ray Bursts (LGRBs) confirmed that these events take place in dense, star-forming environments, joining as an indirect proof the observation of supernovae associated to the bursts and the location in the brightest galaxy regions. Recently, the simultaneous detection of a short Gamma-Ray Burst (SGRB) and a gravitational wave signal occurred, strongly supporting the hypothesis that SGRBs originate instead from the merger of compact objects. The different predictions of the two progenitor scenarios for short and long GRBs should be reflected in a difference in the amount of absorbing matter between the two populations, with SGRBs occurring in less dense environments. Previous studies found that the two column density distributions were indistinguishable when compared in the same redshift range. The samples, though, were relatively small (10-12 SGRBs), and spanned a redshift range $z\lesssim1$.}
   {We update a flux-limited complete sample of \textit{Swift}-based SGRBs (SBAT4, D'Avanzo et al. 2014), bringing it to 25 events and doubling its previous redshift range. We then evaluate the column densities of the events in the updated sample, in order to compare them with the $N_{H}$ distribution of LGRBs, using the sample BAT6ext (Arcodia et al. 2016).}
   {We rely on Monte Carlo simulations of the two populations and compare the computed $N_{H}$ distributions with a two sample Kolmogorov Smirnov (K-S) test. We then study how the K-S probability varies with respect to the redshift range we consider.}
   {We find that the K-S probability keeps decreasing as redshift increases until at $z\sim1.8$ the probability that short and long GRBs come from the same parent distribution drops below $1\%$. This testifies for an observational difference among the two populations. This difference may be due to the presence of highly absorbed LGRBs above $z\sim1.3$, which have not been observed in the SGRB sample yet, although this may be due to our inability to detect them, or to the relatively small sample size.}
   {}


   \maketitle
   \titlerunning{X-ray absorbing column densities of a complete sample of short Gamma Ray Bursts}
   \authorrunning{L. Asquini et al.}
%
%________________________________________________________________

\section{Introduction}

Gamma-Ray Bursts (GRBs) are luminous explosions occurring at cosmological distances that release a huge amount of high-energy photons within a short time. The bimodal distribution of their duration and spectral hardness led to the currently-used classification of long-soft GRBs (LGRBs), lasting more than $\sim2$ s, and short-hard GRBs (SGRBs), which last less than $\sim2$ s \citep{kouveliotou1993identification}.

These differences likely reflect the different origin of these events. Indeed, observational evidence allowed the association of LGRBs to core-collapse supernovae \citep[SNe; supporting the so-called "collapsar model", see][for recent reviews]{hjorth2012chapter,cano2017observer}, while the recent detection of a gravitational wave (GW) source with a simultaneous SGRB seems to be the long-sought "smoking gun" that sets mergers of double neutron stars (NS-NS) as the progenitors of these events \citep{abbott2017multi, goldstein2017ordinary, savchenko2017integral}. According to this scenario, called the compact object binary merger, which includes also neutron star-black hole (NS-BH) binary systems \citep{eichler1989nucleosynthesis, nakar2007short}, the merging objects originate either from (i) a "primordial" binary \citep{narayan1992gamma}, whose component stars were gravitationally bound since their birth, or (ii) a "dynamical binary", formed by means of dynamical capture and possibly exchange in dense stellar environments (e.g. globular clusters) during their relaxation \citep{grindlay2006short, salvaterra2008short}. In case (i) the system must survive the SN explosions of its components, whose natal kick may drive the binary away from the star-forming region. A "fast-merging" variation of the primordial binary channel has been proposed \citep{belczynski2001new, perna2002short, belczynski2006study} and predicts the coalescence to occur in a relatively short time-scale ($\sim10^{7}$ yr), meaning that the bursts would likely occur within their formation sites, possibly in dense star-forming regions. This was motivated by the discovery of the double radio pulsar PSR J0737-3039 \citep{burgay2003increased}, with a merging time of $\sim85$ Myr.
 
The merger model and its evolutionary channels predict other features, besides the concurrent GW emission, that can be observationally tested in order to indirectly probe the nature of SGRB progenitors. One viable way is the study of the environment where SGRBs occur, in comparison to that of LGRBs. According to the collapsar model, LGRBs occur in their star-forming regions, i.e. in an ambient that is quite dense of gas and dust \citep{galama2001high, watson2007, campana2007, schady2011, heintz2018}. 
%%%% AGGIUNGERE QUESTI !!!!!????
For SGRBs, instead, the merger model predicts both dense environments (for systems evolving through the "fast-merging" channel) or less dense regions, such as the outskirts of the host galaxies (for primordial binaries with long merging time that are subject to a natal kick), or for systems that are dynamically formed \citep{grindlay2006short, salvaterra2008short}. 
%%%% AGGIUNGERE QUESTI !!!!!????
The fact that a non-negligible fraction\footnote{Mergers should be kicked out of the star-forming region, but also towards the centre of the host galaxy or behind it along our line of sight, possibly increasing the overall column density.} of the mergers should occur far from star forming regions, and generally at larger offsets from their host galaxies \citep{wang2018possible}, should in principle produce an appreciable difference in the absorption of the optical and X-ray afterglows for the SGRB and LGRB populations.
%{\bf unless} the overall absorption within the host galaxy  {\bf but far from the GRB site} plays a minor role). 
Measuring the total amount of matter needed to produce a given absorption in the afterglow can hence be a viable test to probe the "typical" environments of the two populations, and even discriminate among the different evolutionary channels of SGRBs. This piece of information is, however, hard to obtain from optical spectra because of photoionisation of the material surrounding the burst, and the amount of hydrogen along the line of sight ($N_{HI}$) can only be measured for GRBs at $z\gtrsim 2$. Thus, it is more convenient to work in the X-ray band, where metals are the main absorbers and the measure is less sensitive to photoionisation. 

Such studies have been conducted on many differently-selected samples of LGRBs \citep{campana2010x, campana2012x} by evaluating their intrinsic X-ray absorbing column densities ($N_{H}$), and the high values of $N_{H}$ that were found are consistent with the collapsar model. For SGRBs, \cite{kopavc2012environment} and \cite{margutti2012prompt} independently found that their column density distribution is indistinguishable from that of LGRBs when the two are compared within the same redshift bin. These studies were based on 10-12 events which had a redshift association (which is a mandatory report to derive the intrinsic column density values) spanning up to $z\sim1$, and used them to represent their whole \textit{Neil Gehrels Swift Observatory} (\textit{Swift})-based SGRB samples, which were much bigger (50-60 events) and lacked any information about $z$ for about $75\%$ of them.
 
\cite{d2014complete} tried to overcome these limitations by building a sample that was more representative of the whole SGRB population, namely the SBAT4. They first selected all of the events with a \textit{Swift}/BAT detection and a prompt \textit{Swift}/XRT follow-up (but no afterglow detection is required, so that no X-ray-selection bias was introduced). Then, they restricted the sample to those SGRBs with a prompt emission that had a peak photon flux $P\gtrsim3.5$ ph s$^{-1}$cm$^{-2}$, computed with \textit{Swift}/BAT lightcurve bin width of $64$ ms (making SBAT4 a flux-limited complete sample, note that peak flux threshold and time during which the emission peak is computed are different for SBAT4 and BAT6). Then they required that the events were observed in favorable conditions for redshift determination ($A_{V}<0.5$ mag). This condition implies that the SBAT4 ($16$ events) has a high redshift completeness ($69\%$). These criteria were the analogous\footnote{To select BAT6 and BAT6ext, a peak photon flux $P\gtrsim2.6$ ph s$^{-1}$cm$^{-2}$ in the prompt emission was required. The light curve was binned at 1 s resolution.} of those used by \cite{salvaterra2012complete} to select the BAT6, which is a complete sample of LGRBs ($58$ events with a redshift completeness of $95\%$) that was later updated to BAT6ext by \cite{pescalli2016rate} ($99$ LGRBs with a redshift completeness of $\sim82\%$). 
 
In order to cope with low statistic and redshift incompleteness, in this paper we take a different approach. We first update the SBAT4 peak-flux-limited sample up to April 2016, bringing it to 25 GRBs and extending its maximum redshift from $z\sim1$ to $z\sim2.2$, even though the redshift completeness is lowered to $52\%$. For the SGRBs of the SBAT4 with known redshift, we evaluate the intrinsic column density, while for the rest of the sample we work out the column density in excess of the Galactic value at zero redshift (Section \ref{2}). The 'darkness' of short GRBs of the SBAT4 sample is worked out in Section \ref{3}.
We then make use of Monte Carlo simulations to simulate the intrinsic column density distribution of SGRBs and compare it, on a statistical basis, with the $N_{H}(z)$ distribution of LGRBs (Section \ref{4}). Discussion and calculations are outlined in Section \ref{5}, while conclusions are drawn in Section \ref{6}.  

\section{Column density evaluation}
\label{2}
The total column density can be considered as the summed effect of three main absorbers, i.e. our Galaxy, the intergalactic medium (IGM), and the host galaxy of the GRB. Here we neglect the effects of the IGM \citep{arcodia2018x}, 
%%%% AGGIUNGERE QUESTO!!!!!????
which are negligible especially given the low redshift of the SBAT4 GRBs, and, more importantly, since these effects are the same for GRBs in the SBAT4 and BAT6ext samples. 

Column densities must be evaluated on X-ray spectra that are relative to time intervals where the $0.3-1.5/1.5-10$ keV hardness ratio is constant, in order to avoid unphysical biases in absorption due to spectral changes. We selected time intervals from which to extract the spectra using the lightcurves of the events in SBAT4 (the lightcurves and the spectra were retrieved from the UK Swift Science Data Centre\footnote{http://www.swift.ac.uk/} \citep{evans2009methods}, i.e. the \textit{Swift}/XRT lightcurve and spectra repository). Since we had to avoid the epochs of the lightcurves which presented strong spectral variability, we usually skipped the early times of the afterglow. As a consequence, we selected our data mostly in photon counting (PC) mode.
There were spectra, though, that had too few photons at late times and this prevented any reliable analysis. In these cases, we considered also data from the window timing (WT) mode, early in the afterglow lightcurve. The spectra we worked with were binned with at least one photon in each spectral bin in order to use the C-statistic for fitting \citep{cash1979parameter}. We used the \textmyfont{XSPEC 12.6.1} software \citep{arnaud1996astronomical}, using abundances from \cite{wilms2000absorption} and cross-sections from \cite{verner1996atomic}. We modeled the spectra with a combination of a power law model (\textmyfont{POW}) and two absorption components, one from our Galaxy (\textmyfont{TBABS}, frozen), whose values (taken from the UK Swift Science Data Centre) are those from \cite{willingale2013calibration}, and one at the GRB redshift (\textmyfont{ZTBABS}, which we left free to vary). For those GRBs with unknown redshift, we set $z=0$.\\
For most SGRBs we took data from PC mode and fit them with the model described above. Whenever the number of photons in the selected time slice was not high enough for the fit to return a measurement (i.e. the fit returned an upper limit at $90\%$ confidence level), we searched for other time intervals in the late-time spectrum (PC mode) that presented a constant hardness ratio, in order to increase the statistics and allow for a better fit. We found one suitable interval only for GRB 140903A. For all of the remaining bursts whose analysis resulted in an upper limit, we then looked for an appropriate time interval in the early-time lightcurve (WT data) and found it for 5 SGRBs (3 of which had a redshift association). Further, for GRB 090515 only WT data were available. \\
The underlying assumption that drove the analysis of the GRBs that had two different epochs selected was the fact that, while spectra evolve in time, the quantity of absorbing matter is constant throughout the whole duration of the afterglow. The spectra of the two different epochs of GRB 140903A were simultaneously fit with different power laws but with the same column density value for both epochs. \\
Also, the 5 events that involved WT observations had their two spectra simultaneously fit, but they required a different modeling. As demonstrated by \cite{butler2007x}, early-time spectra cannot be fit with a simple power law, because they might rapidly change. To overcome this issue, the simultaneous fit must be carried out using a simple power law to model PC data and adding a cutoff-energy parameter (left free to vary) to the power law used to fit WT data \citep{campana2010x}. Even if the cutoff-energy, in the end, resulted outside the XRT energy band (i.e. the cutoff is unnecessary) it is important to have this additional freedom to cope with cases in which the X-ray spectrum is instead curved. GRB 090515 was fit with the cutoff power law only.

\begin{figure}
   \centering
   \includegraphics[width=\hsize]{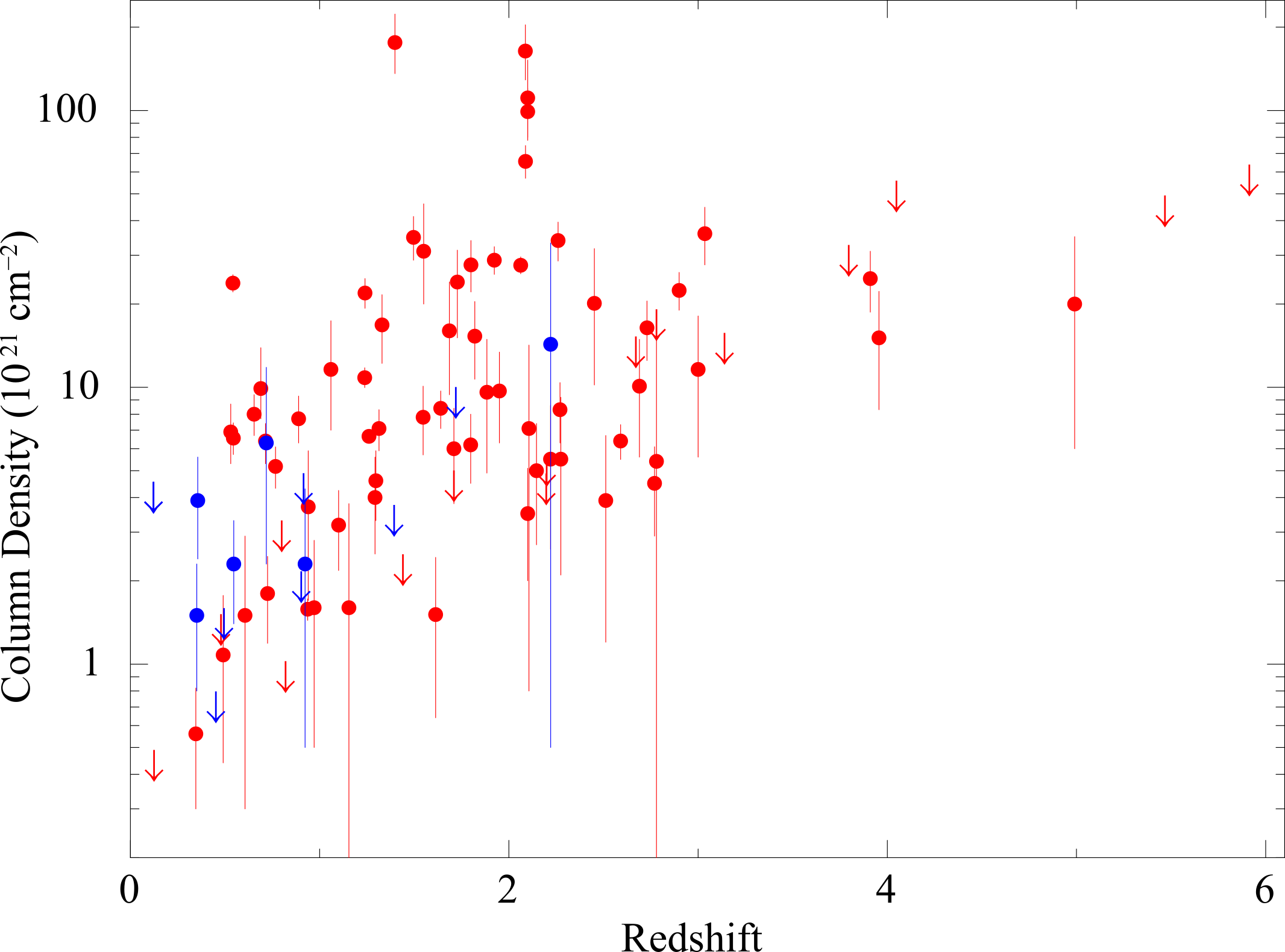}
      \caption{Column density values of the SBAT4 (blue) and the values of the BAT6ext (red) from \cite{arcodia2016dependence}. Downward arrows represent upper limits for both the populations. Although the statistics is very low for SGRBs, the distribution of $N_{H}$ of the events in SBAT4 seems to track that of BAT6ext until $z\sim 1.3$, where LGRBs are slightly more absorbed, and some heavily absorbed LGRBs fill the top region of the plot.}
    \label{shortlong}
   \end{figure}

Of the 13 events with known redshift within the SBAT4, we could compute the intrinsic column density for 6 of them, while for 7 we derived only upper limits. The distribution of the 6 measurements for this sample has a mean value of log$(N_{H}(z)/$cm$^{-2})=21.4$ and $\sigma_{log(N_{H}(z)/cm^{-2})}=0.4$ (which is the same value found by \cite{margutti2012prompt} and \cite{kopavc2012environment}). Of the remaining 12 SGRBs without redshift, we could derive 3 measurements and 9 upper limits, resulting in a total of $9$ measurements and $16$ upper limits for the whole SBAT4. The values of the derived $N_{H}$ are shown in Table \ref{reds}, and they are plotted against redshift in Fig. \ref{shortlong}.  

\begin{table*} 
\small
\centering
\caption{Table of computed column densities for the SGRBs of the SBAT4 sample. Upper limits are marked with an "UL" superscript at the end of the name. Errors and upper limits are given at the 90\% confidence level. Listed in this table there are the GRB name, their redshift and the T$_{90}$ of the prompt emission. The time-slice is the selected interval on which the X-ray spectra were worked out, with the mode noted in brackets. $N_{H}^{Gal}$ is the column density of our Galaxy along the line of sight of the GRB, and its value is from \cite{willingale2013calibration}. $\Gamma$ is the photon index, with the computed (when needed) cutoff-energy in brackets. $N_{H}(z)$ is the intrinsic column density; here we report also the best fit values of those events that were consistent with $0$ within the $90\%$ confidence level (while for some events the fit could not return such a value). All the upper limit values are listed in the relative column, while the last column is the C-stat of the fit, with degrees of freedom in brackets.}
\label{reds}
\begin{tabular}{lccccccccr} 
\hline\hline
GRB & $z$ & $T_{90}$ & Time-Slice (Mode)& $N_{H}^{Gal}$ & $\Gamma$(Cutoff)& $N_{H}(z)$ & UL value& C-stat  \\
  &   & (s)  & (s)  & ($10^{20}$cm$^{-2}$) &$(-)$ (keV) & \multicolumn{2}{c}{($10^{21}$cm$^{-2}$)} & (dof) \\
\hline
$051221A$ &  $0.546$ & $1.4$ & $300-1.5\times10^{5}$ (PC)& $7.52$ &$2.0\pm0.1$ & $2.3^{+1.0}_{-0.9}$ & $-$ & $353.77 (357)$ \\
$060313^{UL}$ &  $-$ & $0.7$ & $4100-93\times10^{4}$ (PC) & $6.17$ &$2.0\pm0.2$ & $0.3^{+0.4}_{-0.3}$&$0.7$  &$292.56 (304)$  \\
$061201^{UL}$ &  $-$ & $0.8$ & $300-800$ (PC)& $6.57$ &$1.6\pm0.2$ & $0.4^{+0.9}_{-0.4}$&$1.3$ & $155.78 (178)$  \\
$070714$B &  $0.923$ & $64$ & $400-1700$ (PC)& $9.83$ &$1.9\pm0.1$ & $2.3^{+2.0}_{-1.8}$ & $-$ & $225.12 (286)$ \\
$080123^{UL}$ & $0.495$  &$115$& $118-163$ (WT) & $2.51$ & $1.1\pm0.3$ ($5.4$) & $0.6^{+0.8}_{-0.6}$ &$1.4$  & $472.92 (541)$ \\
 & &  & $250-2\times10^{4}$ (PC) & &$2.1\pm0.3$ ($-$) & & & \\
$080503^{UL}$ & $-$  &$170$& $150-200$ (WT) & $6.98$ & $0.7\pm0.2$ ($2.9$) & $0$&$0.1$ & $478.32 (591)$ \\
 & &  & $280-1600$ (PC) & &$2.4\pm0.2$ ($-$)& & & \\
$080905$A$^{UL}$ &  $0.122$ & $0.4$ & $400-2000$ (PC)& $13.4$ &$1.6\pm0.3$ & $1.7^{+2.3}_{-1.7}$ &$4.0$  & $106.82 (125)$ \\
$090510^{UL}$ & $0.903$ &$0.3$ & $450-2000$ (PC)& $1.77$ & $1.7\pm0.1$ & $0.8^{+1.1}_{-0.8}$&$1.9$ & $350.17 (326)$ \\
$090515^{UL}$ & $-$  &$0.036$& $70-276$ (WT)  & $2.07$ & $1.4^{+0.3}_{-0.1}$ ($6.4$) & $-$ & $0.2$  &$367.26(381)$ \\
$100117$A$^{UL}$ & $0.915$  &$0.3$& $105-155$ (WT) & $2.91$ & $0.9\pm0.6$ ($4.9$) & $1.8^{+2.5}_{-1.8}$&$4.3$ & $228.76 (331)$ \\
 & &  & $300-600$ (PC) & &$2.3\pm0.3$ ($-$) & & & \\
$100625$A$^{UL}$ & $0.452$ &$0.33$ & $100-708$ (PC)& $2.23$ & $1.4\pm0.2$ & $0$&$0.7$ & $58.56 (63)$ \\
$101219$A & $0.718$ &$0.6$ & $80-199$ (PC)& $5.79$ & $1.4\pm0.3$ &  $6.3^{+5.5}_{-4.0}$ & $-$ &$101.98 (140)$ \\
$111117$A & $2.221$ &$0.47$ & $200-1300$ (PC)& $4.14$ & $1.8\pm0.3$ & $14.3^{+18.9}_{-13.8}$ & $-$ &$82.89 (111)$ \\
$130515$A$^{UL}$ & $-$ &$0.29$ & $80-1.8\times10^{4}$ (PC) & $7.38$ & $1.7^{+0.6}_{-0.5}$ & $0$&$1.4$  & $32.69 (31)$ \\
$130603$B & $0.356$ &$0.18$ & $5000-6500$ (PC)& $2.1$ & $1.8\pm0.2$ & $3.9^{+1.7}_{-1.5}$ & $-$ & $148.60 (206)$ \\
$140622$A$^{UL}$ & $-$ &$0.13$ & $100-9.2\times10^{4}$ (PC) & $5.47$ &$1.6^{+0.4}_{-0.02}$ & $0.2^{+2.8}_{-0.2}$ &$3.0$  & $44.25 (44)$ \\
$140903$A & $0.351$ &$0.30$ & $80-1500$ (PC)& $3.26$ &$1.6\pm0.1$ & $1.5^{+0.8}_{-0.7}$ & $-$ &$444.25 (449)$ \\
 & & & $5000-1.7\times10^{5}$ (PC)& &$1.7\pm0.2$ & & & \\ 
$140930$B & $-$ &$0.84$ & $370-2000$ (PC)& $3.45$ & $2.0\pm0.2$ & $0.6^{+0.6}_{-0.5}$ & $-$ & $148.90(197)$ \\
$141212$A$^{UL}$ & $-$ &$0.30$ & $72-1.9\times10^{4}$ (PC) & $10.3$ &$1.9^{+0.7}_{-0.4}$ & $0$&$1.9$ & $25.64 (31)$ \\
$150423$A$^{UL}$ & $1.394$ &$0.22$ & $80-1200$ (PC)& $1.77$ & $1.4\pm0.2$ & $0$&$3.3$ & $75.43 (128)$ \\
$150424$A & $-$ &$91$ & $4200-10^{4}$ (PC) & $6.02$ &$2.1\pm0.3$ & $1.1^{+1.0}_{-0.8}$ & $-$& $157.90 (181)$ \\
$150831$A$^{UL}$ & $-$  &$11.5$& $116-150$ (WT) & $1.14$ & $1.1^{+0.3}_{-0.6}$ ($25.7$) & $0$&$0.4$ & $218.37 (268)$ \\
 & &  & $201-1300$ (PC) & &$1.7\pm0.3$ ($-$) & & & \\
$151229$A & $-$ &$1.78$ & $4000-4.7\times10^{4}$ (PC)& $2.71$ & $2.1\pm0.2$ & $6.7^{+1.2}_{-1.1}$ & $-$ &$324.82(383)$ \\
$160408$A$^{UL}$ & $-$ &$0.32$ & $100-1400$ (PC) & $4.18$ & $1.6^{+0.6}_{-0.9}$ & $0.3^{+1.2}_{-0.3}$& $1.5$  &$101.43 (109)$ \\
$160410$A$^{UL}$ & $1.72$  &$8.2$& $133-179$ (WT) & $1.77$ & $1.2^{+0.2}_{-0.4}$ ($28.1$) & $0$&$8.8$ &$238.51 (301)$ \\
 & &  & $4200-10^{4}$ (PC) & &$1.6\pm0.3$ ($-$)& & & \\
\hline
\end{tabular}
\end{table*}

\section{Darkness for Short gamma-ray bursts}
\label{3}
Given a large number of upper limits for the absorbing column density or the lack of redshift we tried to derive indirect information on the possibility that SGRBs are absorbed 
by studying their  "darkness''. Darkness has been introduced by Jakobsson et al. (2004; see also Fynbo et al. 2001) to settle on statistical grounds the lack of optical afterglow for a sizeable fraction of 
well localised LGBRs. Three scenarios were put forward involving either obscuration (optical emission is absorbed), high redshift (optical emission is suppressed by damped L$\alpha$ absorbers), 
or low-density ambient medium  (i.e. intrinsically faint optical emission). 
This last possibility occurs when the cooling frequency lies below the X-ray domain, so that the X-ray emission is independent on the circumburst medium but the optical emission depends on the ambient density $n^{1/2}$ \citep{Sari_1999}. Fong et al. (2015) showed that this occurs for almost half of SGRBs in their sample.
In order to assess if a GRB was not detected, a link to the overall afterglow properties is important. This can be obtained by computing 
the  optical-to-X-ray spectral index at a given time. 
In the fireball model, the spectral index $\beta$ (with $F_{\nu}\propto \nu^{-\beta}$) for connecting X-ray ($\sim 10^{18}$ Hz) and optical ($\sim 10^{14}$ Hz) frequencies, $\beta_{\rm ox}$,
is expected to lie in the 0.5-1.25 range, unless the optical emission is dimmer for one of the reasons described above; GRBs whose afterglows has $\beta_{\rm ox} < 0.5$ are defined as "dark"  (Jakobsson et al. 2004, see also van der Horst et al. 2009).

\begin{figure} [h!]
   \centering
   \includegraphics[width=\hsize]{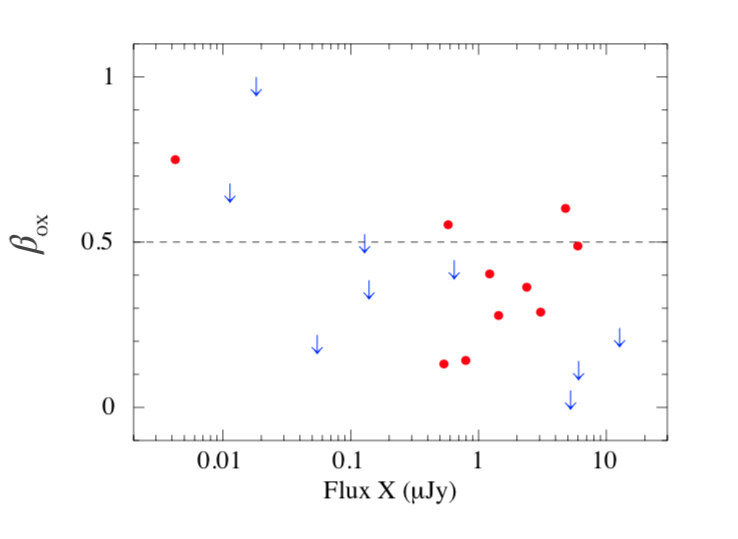}
   \includegraphics[width=\hsize]{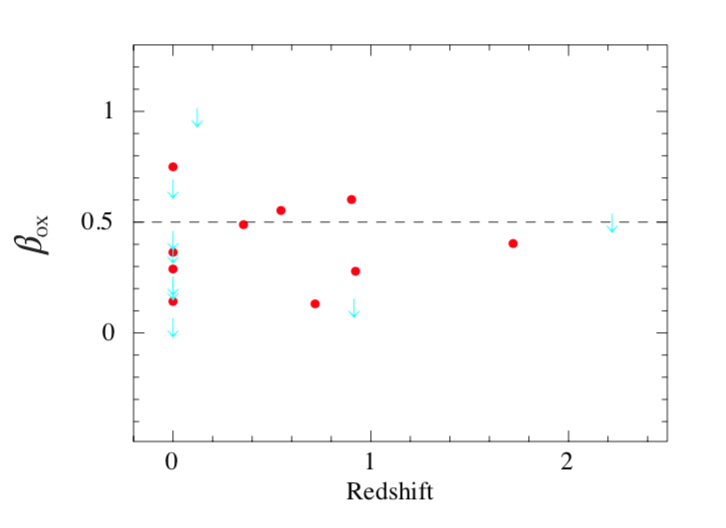}
   \includegraphics[width=8.3cm]{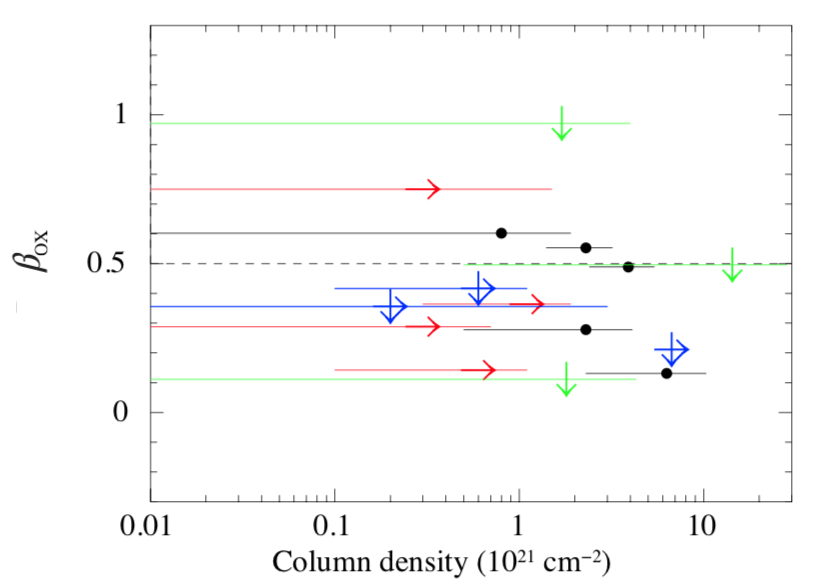}
     \caption{Upper panel: $\beta_{OX}$ for the short GRBs of the SBAT4 sample. Upper limits are shown with a downward arrow. The horizontal line marks the limit of GRB darkness.
     Medium panel: beta$_{OX}$ as a function of redshift. GRBs with unknown redshift are shown at $z=0$. Lower panel: $\beta_{OX}$ as a function of the intrinsic column density. Downward arrows indicate upper limits on $\beta$. Rightward arrows indicate GRBs with unknown redshift for which the column density has been computed at $z=0$.}
    \label{darkfig}
   \end{figure}

The spectral index $\beta_{\rm ox}$ for LGRBs was evaluated at 11 hr after the onset (observer frame). 
This is hardly feasible for SGRBs, since their emission is already gone beyond the reach of optical and X-ray facilities at that time.
We evaluate $\beta_{\rm ox}$ for our complete sample of SGRBs at 1 hr. Even at such a close time to the prompt, a number of SGRBs do not show a detection.
For the X-ray fluxes, we relied on Swift observation with the XRT. Data were taken from XRT web pages\footnote{http://www.swift.ac.uk//burst\_analyser/} (Evans et al. 2009). 
Data were interpolated and in a few cases extrapolated using a fit of the entire (power law) light curve.
For the optical fluxes, we relied on data found in the literature (for SGRBs up to 2014 on \cite{fong2015decade}); on papers and GCN circulars for the others). 
For optical data, we consider $r$ band and usually we 
derived our values thanks to a back-extrapolation of the light curve including early time upper limits when available (the earliest point in the back extrapolated light curves are in the 2-6 hr range). In one case  we forward-extrapolated from 30 min. For 9 SGRBs we selected the upper limit closer in time to 1 hr (from 30 min to 2 hr).
For 6 SGRBs we were not able even to place a meaningful upper limit on the $r$-band optical flux at 1 hr due to the lack of any data or to the presence of late ($\sim 12$ hr) upper limits.
Our findings are summarised in Fig. \ref{darkfig}, where we show the spectral parameter 
$\beta_{\rm ox}$ as a function of the X-ray flux, of redshift, and of the intrinsic column density. 
As it might have been expected, a number of short GRBs are dark ($\sim 75\%$). This is at variance with long GRBs, where a fraction of $30\%$  has been found as dark \citep{melandri2012dark}. No clear trend with redshift or intrinsic column density is immediately apparent from our analysis (see Fig. \ref{darkfig}), therefore suggesting that the tenuous ambient medium is the root source for faintness of the optical emission, in most of the cases.
In our sample, there is just one potential heavily absorbed SGRB: GRB151229, with a column density at $z=0$ of $6.7^{+1.2}_{-1.1}\times 10^{21}$ cm$^{-2}$, which will overcome the threshold  of $10^{22}$ cm$^{-2}$ for 
redshifts $z\gsim 0.2$ and  of $10^{23}$ cm$^{-2}$ for redshifts $z\gsim 2$.

\section{Simulations and comparison with BAT6ext}
\label{4}
The lack of redshift in the sample ($13/25$, $52\%$ completeness) and the even smaller number of measured column densities for bursts with associated \textit{z} we got ($6/25$), make the distribution of $N_{H}(z)$ we derived, also including upper limits, not fully representative of the whole population of SGRBs.
The lack of redshift, and therefore of a proper $N_{H}(z)$ evaluation for almost half of the SBAT4 sample, is particularly restrictive for our studies, hampering \textit{ab initio} any survival analysis of the two GRB populations (long and short GRBs). We decided then to rely on Monte Carlo simulations to cope with redshift incompleteness and upper limits. 

The method proceeded as follows. For GRBs with measured $N_{H}$, the simulation extracted the column density values from a Gaussian distribution that peaked on the measured value, assuming a symmetric error that is the largest of the two reported in Table 1 (a lower limit was fixed in the Gaussian to prevent the occurrence of unphysical negative values of $N_{H}$). If the redshift of the GRB was unknown, we put the simulated value at zero redshift ($N_{H}(0)$). If needed, a value of z was randomly assigned, based on the observed redshift distribution of the SBAT4 sample as updated in this work, i.e. assuming it as valid also for the fraction of the sample without a redshift measurement. We then obtained the value $N_{H}(z)$ by multiplying $N_{H}(0)$ by the scaling factor $(1+z)^{2.4}$ \citep{campana2014effective}.

To check for the effectiveness of the choice of this redshift distribution, we compared  the column density distribution of short GRBs with and without redshift, evaluated setting all of the events at $z=0$. We worked out a Monte Carlo simulation to provide 1,000 realisations of our sample to deal with upper limits and errors. We split the simulated column densities into two sub-samples depending if the redshift was available or not, and compared the two by means of a Kolmogorov-Smirnoff (K-S) test. The mean probability over the 1,000 simulations is  $56\%$ (with a standard deviation of $27\%$). This testifies that the use of the observed redshift distribution as a reasonable assumption.
%a value of \textit{z} was randomly assigned, based on the observed redshift distribution of the SBAT4 sample \citep{ghirlanda2016short}.
%We then obtained the value $N_{H}(z)$ by multiplying $N_{H}(0)$ by the scaling factor $(1+z)^{2.4}$ \citep{campana2014effective}. {\bf To support this choice, we compared the column density distribution of short GRBs with and without redshift, evaluated for all of them at $z=0$. We worked out a Monte Carlo simulation to provide 1,000 realisations of our sample to deal with upper limits and errors. We split the simulated column densities into two sub-samples depending if the redshift was available or not, and compared the two by means of a Kolmogorov-Smirnoff (K-S) test. The mean probability over the 1,000 simulations is  $56\%$ (with a standard deviation of $27\%$). This justifies our use of the observed redshift distribution.}
 %If the measured $N_{H}$ was an upper-limit, a finite upper threshold was fixed for the extraction of values. 
 
After simulating 1,000 times the sample of SGRBs (with the randomly assigned redshift, if needed), the same procedure was applied to the sample of $99$ LGRBs of the BAT6ext, with the same caveats on upper limits and redshifts, taking the $N_{H}$ values from \cite{arcodia2016dependence}. We took the redshift distribution from \cite{pescalli2016rate}. This process resulted in 1,000 mock samples of SGRBs and LGRBs each.

\begin{figure} 
   \centering
   \includegraphics[width=\hsize]{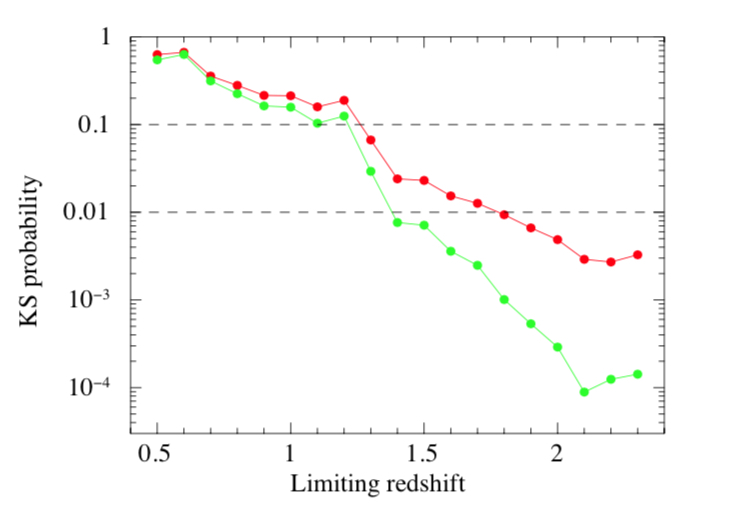}
      \caption{K-S probability between the two populations of LGRBs in the BAT6ext (from \cite{arcodia2016dependence}) and the SBAT4, as a function of the applied cut in $z$. Red dots represent the median value of the obtained K-S probability distribution, while green dots are their logarithmic mean value. As the redshift range increases, the probability follows a decreasing trend. The two populations are likely drawn from the same distribution up to $z\sim 1.2$, then the probability crosses a band (between the green dashed lines, corresponding to a probability of $1\%$ to $10\%$) where the K-S test is ambiguous, and keeps decreasing. At $z\sim1.4$ the logarithmic mean of the K-S probability has dropped below the $1\%$ threshold, and at $z\sim1.8$ also the median value indicates that the two populations are likely different.}
    \label{ks}
   \end{figure}

We then compared the two simulated populations, each set at a time, through a two-sample Kolmogorov-Smirnoff test. 
Since the redshift range of LGRBs in the BAT6ext is much broader than that of SGRBs in SBAT4 ($z_{max}\sim5.9$ for LGRBs vs. $z_{max}\sim2.2$ for SGRBs), we cut the BAT6ext simulated $N_{H}$ distribution at $z\sim 2.3$, in order to compare the two distributions in the same redshift bin. The two-sample K-S resulted in a logarithmic mean probability of $1.4\times10^{-4}$ and a median of $3.3\times10^{-3}$.\\
We then cut both the SBAT4 and BAT6ext distributions from $z=0.5$ to $z=2.3$ in steps of $0.1$, filling the plot in Fig. \ref{ks}, in order to test how the probability varied as a function of the redshift cut.  

We also split the two datasets into two different redshift bins 0-0.7 and 0.7-2.3, and carried out the K-S test separately (this is because the two distributions populate the redshift span differently and we wanted to test if this difference has an effect). The results confirm our findings: the two distributions within the 0-0.7 redshift bin are comparable whereas in the 0.7-2.3 redshift bin are different at more than $3\,\sigma$ level.
Finally, we added by hand a heavily absorbed SGRBs to the sample, with $N_{H}(z)=10^{23}$ cm$^{-2}$ at $z=1.5$. We then run again the entire Monte Carlo simulation. The inclusion of this single heavily absorbed SGRB weakens considerably the K-S results with a median overall probability of $\sim 10^{-2}$.

As a further check, we also repeated the simulations using the redshift distribution presented in \cite{ghirlanda2016short}, that is representative of the whole SGRB population, finding a logarithmic mean probability of $2.5\times 10^{-3}$ and a median of $3.0\times10^{-3}$, in agreement with the results reported above.

\section{Discussion}
\label{5}
%The results of these simulations are consistent with the conclusions drawn in previous works. Indeed, the two populations (SGRBs and LGRBs) are found to be unquestionably different when considered without any restriction in redshift, but this difference vanishes when considering only the closest events ($z\lesssim1.3$).
Previous works found that the populations of LGRBs and SGRBs are different when considered without any restriction in redshift, but consistent when compared in the $z\sim0-1$ range, that was the range of their SGRB samples \citep{kopavc2012environment,margutti2012prompt, d2014complete}. The consistency of the two distributions of $N_{H}$ up to $z\sim1$ was interpreted as evidence that the majority of SGRB progenitors evolve via the "fast-merging" channel, i.e. they share a similar environment with LGRBs, with the caution that column densities, being integrated quantities, are a good proxy for the host-galaxy properties and not of the circumburst medium. Under this hypothesis, one would expect that the extension of the sample to $z\sim2.2$ would confirm the consistency of the two distributions of $N_{H}$. Instead, we found that the two populations become more and more distinguishable when compared in a redshift range that is above the $z\sim1.3$ threshold, with SGRBs being less absorbed.
This may be due to the fact that above $z\sim1$ the plot in Fig. \ref{shortlong} is populated by highly absorbed LGRBs. Although the definition of "dark" LGRBs is not based on X-ray absorption (see Section \ref{3}), \cite{campana2012x},  \cite{fynbo2001detection}, \cite{jakobsson2004swift} and \cite{van2009optical} showed that there is a strong correlation between the darkness of these events and their high $N_{H}$ value and that they are likely due to absorption occurring in the circumburst medium. No SGRB with such high $N_{H}$ $\sim10^{23}$ cm$^{-2}$ is observed at the same redshift (between $z\sim1.3$ and $z\sim2.1$), as one would expect if the two environments were substantially different, at least beyond $z\sim1.3$. We note, however, that the lack of heavily absorbed SGRB with known redshift might be due to an observational bias. The dense medium where heavily absorbed GRBs occur can suppress the optical afterglow emission, making the optical afterglows of SGRBs occurring in such a dense medium too faint to be detected by current facilities. Besides this would result in less accurate positions (that would be only X-ray based) that would make difficult to securely associate a host galaxy (almost all SGRB redshifts measured so far were obtained from optical spectroscopy of their host galaxies).
  
Redshift-selection effects may indeed play a role in our results, biasing the SBAT4 sample towards less-absorbed (and thus observable) SGRBs at $z\gtrsim1$. If, however, the indication for a difference in absorption that we find were indeed intrinsic, our results cannot discern between the expectations deriving from the primordial binary and the dynamical formation scenarios, since they both predict lower $N_{H}$ values for SGRBs with respect to LGRBs. Indeed, the use of column densities alone cannot discriminate whether such a difference in absorption (which may also be due to different density, metallicity or abundances) is caused by the host galaxies properties  of the two populations \citep{buchner2016galaxy}, by a difference in the sub-galactic environments where the GRBs occur, or by different selection effects. However, both the possibilities of different host galaxy type and/or different environment indirectly suggest that LGRBs and SGRBs do not share the same progenitors.

%__________________________________________________________________

%______________________________________________________________

\section{Conclusions}
\label{6}
The evaluation of the column densities for the population of SGRBs is an indirect tool to probe the consistency of the compact object merger model, alongside the direct proof that is the simultaneous detection of an SGRB and a gravitational wave source \citep{abbott2017multi}. The comparison between the $N_{H}$ distribution of SGRBs and that of LGRBs is supposed to highlight the discrepancy in absorption that is predicted by the different progenitor scenarios for the two populations. Given the typical faintness of SGRB afterglows, this comparison should be carried out on homogeneous samples that reduce any selection bias due to redshift or X-ray afterglow properties. These features are found in the SBAT4 for SGRBs \citep{d2014complete} and in the BAT6ext for LGRBs \citep{salvaterra2012complete, pescalli2016rate}, since they are complete flux-limited samples that well represent the whole bright populations of their respective GRB classes. In this paper we extended the SBAT4 flux-limited sample to 25 events, raising its redshift range from $z\sim1.3$ to $z=2.2$ \citep{selsing2017host}. We then computed the column density values of the $\sim50\%$ of the sample which had a redshift measurement, obtaining 6 detections and 7 upper limits. We used Monte Carlo simulations of the populations of SBAT4 and BAT6ext \citep{salvaterra2012complete, pescalli2016rate, arcodia2016dependence} to overcome the low statistics, approximating the probability distribution for each measurement as a Gaussian and thus assuming a symmetrical error. By using the observed redshift distribution of the SBAT4 sample \citep{d2014complete, ghirlanda2016short} we were able to include also the events that lacked a redshift association. We were hence able to make a comparison between the two $N_{H}$ distributions, which was carried out by a two-sample K-S test. We first studied the whole SBAT4 and BAT6ext in the same redshift bin ($z\sim2.3$), obtaining that the two populations are unlikely drawn from the same distribution. We then compared the two samples applying to both a cut in redshift from $z=0.5$ to $z=2.2$ in steps of $0.1$. Our results up to $z\sim 1.3$ are consistent with previous works, i.e. the two populations are indistinguishable from each other. The K-S probability value continues decreasing, until at $z\sim 1.8$ ($z\sim 1.4$, if we rely on the logarithmic mean value only) the two parent distributions are significantly different (below $1\%$), and become more and more distinct the higher the redshift cut is placed. These results suggest that SGRBs are less absorbed than LGRBs, as one would expect if the environments where they occurred were less dense, and they are in agreement with what is predicted by the merger scenario, both through the primordial binary or the dynamical evolutionary channels. However, this difference in absorption does not emerge firmly until $z\sim 1.8$ ($1.4$ for logarithmic mean only) and is totally absent until $z\sim 1$. This might indicate that the fast merging channel for SGRBs is less effective starting from $z\gsim 1$. 
 Alternatively, this may mirror the fact that the presence of the so-called "dark" LGRBs \citep{campana2012x, fynbo2001detection, jakobsson2004swift, van2009optical} in the redshift range between $z\sim 1.3-2.2$ that may be causing the two distributions to part. This is indeed confirmed by the fake addition of a heavily absorbed SGRB to the simulated sample: the difference between the two populations vanishes. Hence, the current instrumental challenge to detect heavily absorbed SGRB at $z\gtrsim1$ may be biasing the sample towards less absorbed events.
%Further, the use of $N_{H}$ distributions alone does not allow to locate the absorber, so that it is not possible to tell whether the circumburst medium or the host galaxy are accountable for this difference. 
Future updates of the SBAT4, with possibly an increase in redshift completeness and range, may continue these studies, enriching the statistics about these events. 

\begin{acknowledgements}
We thank R. Arcodia for useful discussions. 
 We thank the anonymous referee for useful comments. This work made use of data supplied by the UK Swift Science Data Centre at the University of Leicester.
We also acknowledge support from ASI grant I/004/11/3.
\end{acknowledgements}

%----------------------------------
\bibliographystyle{aa}
\bibliography{biblio.bib}

\end{document}